\documentclass[12pt]{book}

\usepackage[dvips]{graphicx,color}
\usepackage{makeidx,tsukuba}

\makeauthorindex
\makeindex

\begin{document}

\BookTitle{\itshape The 28th International Cosmic Ray Conference}
\CopyRight{\copyright 2003 by Universal Academy Press, Inc.}
\pagenumbering{arabic}

\def\simleq{\; \raise0.3ex\hbox{$<$\kern-0.75em \raise-1.1ex\hbox{$\sim$}}\; }
\def\simgeq{\; \raise0.3ex\hbox{$>$\kern-0.75em \raise-1.1ex\hbox{$\sim$}}\; }
\newcommand{\eV}{{\rm eV}}
\newcommand{\Mpc}{{\rm Mpc}}
\newcommand{\kpc}{{\rm kpc}}
\newcommand{\muG}{\mu{\rm G}}

\chapter{
Constrained Simulations of the Magnetic Field in the Local Supercluster
and the Propagation of UHECR}

\author{ Klaus Dolag$^1$, Dario~Grasso$^2$, Volker~Springel$^3$ and 
Igor Tkachev$^4$ \\
{\it (1) Dipartimento di Astronomia, Universit\`a di Padova, Padua, Italy\\
(2) Scuola Normale Superiore, Pisa - Italy \\
(3) Max-Planck-Institut f\"ur Astrophysik, Garching, Germany\\
(4) CERN - Theory Division, Geneve, Switzerland }\\} 

\section*{Abstract}
Magnetic fields (MF) in the Local Supercluster (LSC) of galaxies may
have profound consequences for the propagation of Ultra High Energy
Cosmic Rays (UHECR). Faraday rotations measurements provide some
informations about MF in compact clusters. However, very few is known
about less dense regions and about the global structure of MF in the
LSC.  In order to get a better knowledge of these fields we are
performing constrained magnetohydrodynamical simulations of the LSC
magnetic field.  We will present the results of our simulation and
discuss their implications for the angular distribution of expected UHECR
deflections.

\section{Introduction}
Several arguments suggest that Ultra High Energy Cosmic Rays (UHECR)
are extragalactic in origin (see e.g.[1]). Propagation of electrically
charged CR primaries is affected by intergalactic MF.  Resulting
picture crucially depends on magnetic field mean intensity and power
spectrum.  MF have been observed in several clusters of galaxies
[7]. An indisputable evidence of these fields is provided by diffuse
cluster-wide synchrotron emission from several clusters compatible
with a field strength larger than $0.1~\muG$. In addition, indirect
observational evidence of the magnetization of the Inter Cluster
Medium (ICM) comes from Faraday Rotation Measurements (RM) of the
polarized radio sources located within or behind the clusters.
Analyzing these observations several authors (see e.g.  [3]) found
that the ICM in clusters are permeated with a high filling factor by
magnetic fields at a level of $1-10~\muG$ with a correlation length of
$10\div 100~\kpc$ extending up to 1 Mpc from the cluster
center. Beyond this distance the IGM density becomes too low to allow
an observable Faraday rotation since the effect is proportional to the
free electron density. In this case only upper bounds are available
which are $B < 10^{-9} \div 10^{-8}$ G in unclustered regions and a
significantly weaker upper limit $B \simleq 10^{-6}$ in the sheets or
the filaments of galaxy clusters [2].  What is most crucial for UHECR
reaching the Earth is the MF in the Local Supercluster (LSC) since it
comprise most of the GZK volume.  The MF structure of the LSC is very
poorly known. RM allowed to determine its intensity only in some
overdense regions like the Virgo cluster where it was found to be
about $10^{-6}~$ G.  A method to determine the power spectrum of IGMF
from a systematic analysis of RM of a large number of distance sources
may allow a better knowledge of the LSCMF in a not too distant future
[5].  Meanwhile, however, the only way to improve our understanding of
UHECR observational data is to estimate the LSCMF large scale
structure by numerical simulations.

\section{MHD simulations in clusters}
Magneto Hydrodynamical (MHD) simulations of MF evolution in galaxy
clusters have been already performed by one of us (KD) and
collaborators [4]. This kind of codes combines the merely
gravitational interaction of a dominant dark-matter (DM) component
with the hydrodynamics of a magnetized gas to simulate the formation
of magnetized galaxy clusters. Initial conditions are specified at
redshift $z_{in} \sim 20$ by a random choice of density perturbations
compatible with standard $\Lambda$CDM cosmology and an initial seed
field to be tuned to reproduce RM of clusters at redshift close to
zero. The MF is amplified by adiabatic compression during cluster
collapse and, more efficiently, by the Kelvin-Helmholtz instability
where merger events give rise to strong shear flows. It was found that
the simulation reproduce quite well the general features of the
observed RM with an initial field strength of $\sim 5 \times 10^{-9}\,
$G.  In the outer regions of the clusters, the average field strength
follows the gas density profile quite well though with a steeper slope
compared to estimates based on flux conservation and adiabatic
compression ($B \propto \rho_{\rm gas}^{2/3}$), showing that shear
flow have to play an important role there.  As a consequence the
required initial intensity of the MF is smaller than expectation based
on a mere adiabatic amplification.  The MHD simulations are designed
to study individual clusters within a cosmological environment, and
therefore can not describe voids or filaments.  Nevertheless an order
of magnitude estimate can be obtained on the basis of the following
simple argument.  In the voids, as well as in low density regions
around the LG, we do expect the adiabatic compression to give a rather
good description of the MF evolution.  Using flux conservation, which
implies $B(z = 0) = B(z_{\rm in}) (1 + z_{\rm in})^{-2}$, we see that
a field strength of $5 \times 10^{-9}$ G at $ z_{\rm in} = 20$, as
required for the MHD simulations to reproduce RM of clusters,
corresponds to the present time intensity of $B(z = 0) \sim 10^{-11}$
G in the unclusterized IGM.  The rms deflection angle of a particle of
charge $Ze$ traveling over the distance $d$ through an irregular MF
with a coherence length $l_c \ll d$ and local intensity $B$ is given
by the expression
\begin{equation}
\theta(E,d) \simeq 1^o\; Z \left(\frac{10^{20}~\eV}{E}\right)
\left(\frac{d}{10~\Mpc}\right)^{1/2}
\left(\frac{l_c}{1~\Mpc}\right)^{1/2}
\left(\frac{B}{10^{-9}~{\rm G}}\right)~.
\label{defle}
\end{equation}
It is clear that a MF with strength $\sim 10^{-11}$ G have a
negligible influence on UHECR propagation.  Therefore we expect
significant deflections only when UHECR trajectory crosses cluster or
a strongly magnetized filament.  In that case deflection of few
degrees can be produced (see Fig.\ref{defle}).  A different situation was
examined recently by Sigl at al. [9].  The authors of [9] performed an
Eulerian hydro+N-body simulation of the entire LSC and added to it a
MF which follows passively the gas. Such simulation, however, is not a
constrained simulation (see below) which implies that it is not suited
to reproduce the real spatial distribution of matter and of magnetic
fields. In [9] the authors chosen the observer position to lie in a
region with a rather strong magnetization $B \sim 0.1 \muG$. This was
required in order to reproduce the observed isotropy in the arrival
directions of UHECR under the hypothesis that the source distribution
trace that of the matter in the LSC. Such an hypothesis, however, has not
been proved so far and other reasonable possibilities have been
discussed in the literature.  In our opinion the assumption of a
widely extended MF with intensity around $10^{-7}$ G in the
surrounding of the LG is unrealistic. Indeed the LG lie in a quite
peripherical region of the LSC and it is neither within a rich cluster
nor within a strongly radio emitting region. So far these are the only
extended systems where observational evidence of MF with strength
comparable or larger than $0.1~\muG$ have been found.

\section{Constrained simulation in the LSC}
Realistic maps of expected UHECR deflections by EGMF as a function of
distance to plasuable sources are highly desirable. Ideally, such maps
should reflect obsreved distribution of galaxies and clusters and
therefore different kind of simulations is needed. A promising
approach is that of {\it constrained simulations}, i.e. simulations
with initial conditions constrained to reproduce the observed
large-scale structures in the nearby universe.  Simulations of this
kind have been already performed for the collisionless DM and the
gaseous components of the LSC (see e.g.[6,8]) which succeed to recover
the main observed structures starting from initial conditions at high
redshift compatible with standard cosmology.

We are now working to produce a fully MHD constrained simulation of
the LSC.  Our code incorporates the MSPH technique [4], which was
developed to follow MF evolution in simulated galaxy clusters, into a
constrained N-body simulation of the DM and the gas in the LSC.
Hopefully, at the end we will have simulated 3D maps resembling real
local structure of MF fields. Once we will be in possession of such
maps we will trace UHECR trajectories through the magnetic web. We
will not limit ourselves by UHECRs sources located inside the LSC.  In
particular, we are interested to find out if and when correlations of
charged UHECR with BL Lacs, found in Ref. [10], are compartible with
expected EGMF structure.  In other words, one of our main goals is to
produce maps of expected UHECR deflections for a wide range of source
distances.

We will present detailed results of our simulation in the course of
ICRC2003.  We provide here a sample of our simulation. Fig. 1
represents a map of deflections of protons with energy $E = 4\times
10^{19}~\eV$ derived from a Virgo like cluster, seen from 5 Mpc.  The
MF intensity at the cluster center corresponds to $\sim 1~\muG$ which is
representative of a clusterized region. Rays have been propagated
about 10-15 Mpc in each direction. Deflections larger than 5 degrees
are cut at this value to increase dynamical range of the plot for
small deflections.

\begin{figure}[t]
  \begin{center}
    \includegraphics[height=16pc]{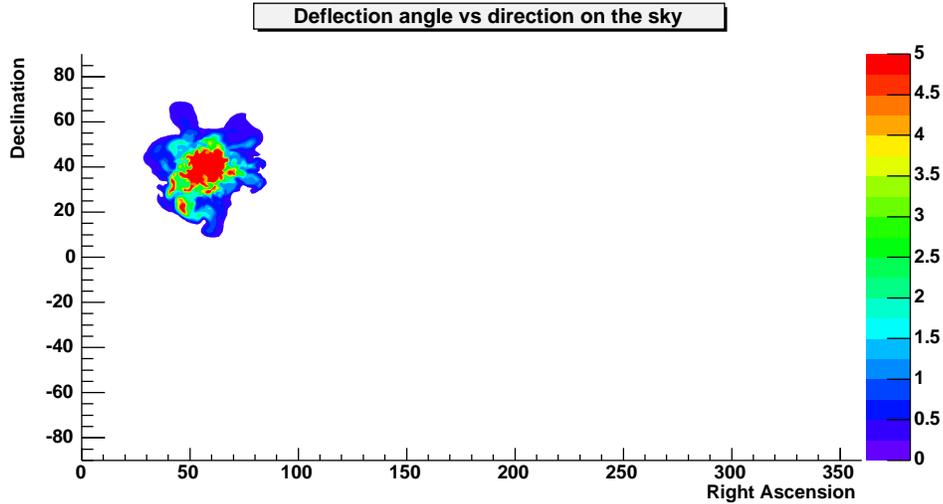}
  \end{center}
  \vspace{-0.5pc}
  \caption{Deflection angle as a function of the direction.}
\end{figure}

\section*{Bibliography}
\noindent
1. Berezinsky~V., Gazizov~A.~Z. and Grigorieva~S.~I.,
arXiv:astro-ph/0210095\\
2. Blasi~P., S.~Burles S. and A.~V.~Olinto~A.V.
1999,  Astrophys.\ J.\  {\bf 514}, L79\\
3. Clarke~T.E., Kronberg~P.P. and Boehringer~H.
2001, Astrop.\ J.\  {\bf 547}, L111\\
4. Dolag~K., Bartelmann~M. and Lesch~H.,
2002, A\&A {\bf387}, 383\\
5. Kolatt~T.,
1998, Astrophys.\ J.\  {\bf 495}, 564\\
6. Kravtsov~A.V., Klypin~A.A. and Hoffman~Y.,
2002, Astrop.J., {\bf 571}, 563\\
7. Kronberg~P.P., 1994, Rep.\ Prog.\ Phys.\ {\bf 57}, 325\\
8. Mathis~H. et al.,
2002, MNRAS, {\bf 333}, 739\\
9. Sigl~G., Miniati~F. and Ensslin~T.,
arXiv:astro-ph/0302388\\
10. Tinyakov~P.G. and Tkachev~I.I.
2002, Astropart.\ Phys.\  {\bf 18}, 165

\endofpaper
\end{document}